\documentclass[twocolumn]{aastex63}
\received{}
\revised{}
\accepted{}
\submitjournal{ApJL}

\shorttitle{Spontaneous Generation of $\delta$-sunspots}
\shortauthors{Toriumi \& Hotta}

\begin{document}

\title{Spontaneous Generation of $\delta$-sunspots in Convective Magnetohydrodynamic Simulation of Magnetic Flux Emergence}

\correspondingauthor{Shin Toriumi}
\email{toriumi.shin@jaxa.jp}

\author[0000-0002-1276-2403]{Shin Toriumi}
\affiliation{Institute of Space and Astronautical Science (ISAS)/Japan Aerospace Exploration Agency (JAXA), 3-1-1 Yoshinodai, Chuo-ku, Sagamihara, Kanagawa 252-5210, Japan}

\author[0000-0002-6312-7944]{Hideyuki Hotta}
\affiliation{Department of Physics, Graduate School of Science, Chiba University, 1-33 Yayoi-cho, Inage-ku, Chiba 263-8522, Japan}



\begin{abstract}
Observations reveal that strong solar flares and coronal mass ejections tend to occur in complex active regions characterized by $\delta$-sunspots, spot rotation, sheared polarity inversion lines (PILs), and magnetic flux ropes. Here we report on the first modeling of spontaneous $\delta$-spot generation as a result of flux emergence from the turbulent convection zone. Utilizing state-of-the-art radiative magnetohydrodynamics code {\it R2D2}, we simulate the emergence of a force-free flux tube in the convection zone that stretches down to $-140\ {\rm Mm}$. Elevated by large-scale convective upflows, the tube appears on the photosphere as two emerging bipoles. The opposite polarities collide against each other due to the subsurface connectivity, and they develop into a pair of closely-packed $\delta$-spots. The Lorentz force drives the spot rotation and a strong counter-streaming flow of $10\ {\rm km\ s}^{-1}$ at the PIL in $\delta$-spots, which, in tandem with local convection, strengthens the horizontal field to 4 kG and builds up a highly-sheared PIL. In the atmosphere above the PIL, a flux rope structure is created. All these processes follow the multi-buoyant segment theory of the $\delta$-spot formation, and they occur as a natural consequence of interaction between magnetic flux and turbulent convection, suggesting that the generation of $\delta$-spots and the resultant flare eruptions may be a stochastically determined process.
\end{abstract}



\section{Introduction} \label{sec:intro}

Solar flares and coronal mass ejections are the plasma process through which magnetic energy is rapidly converted to heat, kinetic energy, and accelerated high-energy particles \citep{2011LRSP....8....6S}. It is known that major flares emanate from {\it complex} active regions \citep{2019LRSP...16....3T}. For instance, statistical studies revealed that the $\delta$-sunspots, in which umbrae of opposite polarities are in close vicinity of each other that they share a common penumbra, are prone to strongest flares \citep{1960AN....285..271K,2000ApJ...540..583S,2017ApJ...834...56T}. Other key features include: spot rotation \citep{2003SoPh..216...79B,2008MNRAS.391.1887Y}; shear flows along the polarity inversion line (PIL) \citep{1976SoPh...47..233H,1982SoPh...79...59K}; PIL with strong magnetic field, $B_{z}$ gradient, and shear \citep{1984SoPh...91..115H,2007ApJ...655L.117S}; magnetic channel \citep{1993Natur.363..426Z,2008ApJ...687..658W}; and magnetic flux rope \citep{2006SSRv..124..131G}.

A major difficulty faced while understanding the formation of flare-productive active regions through emergence from the convection zone is that we cannot investigate the subsurface magnetic field from direct optical observations. Therefore, a number of magnetohydrodynamic (MHD) flux emergence simulations have been conducted in the past decades.

One of the suggested scenarios for $\delta$-spot formation is the multi-segment buoyant model, where a subsurface magnetic flux rises at two locations and appears on the surface as a pair of emerging bipoles. Within this quadrupolar region, the inner polarities of opposite signs collide against each other to form a $\delta$-structure with a sheared PIL in between. This scenario was advocated by \citet{2014SoPh..289.3351T}, who modeled the emergence of a single horizontal tube that is initially made buoyant at two segments. They found that the confinement of the opposite polarities on the surface occurs because the two emerging sections are connected by a dipped field beneath the surface. This situation was followed by \citet{2015ApJ...806...79F}, while \citet{oi2007}\footnote{\url{http://www-space.eps.s.u-tokyo.ac.jp/group/yokoyama-lab/thesis/oi2017\_mastereruption.pdf}} and \citet{2019arXiv190901446S} modeled flare eruptions from the quadrupolar system. Other scenarios include the emergence of a kink-unstable flux tube \citep{1998ApJ...505L..59F,2015ApJ...813..112T,2018ApJ...864...89K} and the collision of two emerging tubes \citep{2007A&A...470..709M,2018ApJ...857...83J,2019NatAs...3..160C}.

Although these simulations succeeded in reproducing some key aspects of flaring regions, many were performed in highly idealized or controlled situations. For instance, the convection zone is mimicked as a plane-parallel atmosphere without including convective flows; the tube's emergence is triggered by artificially reducing the density from the tube; or the emerging flux is kinematically advected into the domain through the boundary. However, in reality, the spot formation occurs as a natural consequence of the interaction between magnetic flux and background convection, which has been difficult for modelers to accommodate.

In this Letter, we report on the first spontaneous $\delta$-spot formation that follows the multi-buoyant segment scenario as a result of flux emergence from the turbulent convection zone. Utilizing the newly developed radiative MHD code, which solves thermal convection of various scales from 100 Mm-sized cells to surface granules, we are now able to overcome the above issues and address the effect of turbulence on emerging flux, the $\delta$-spot formation, and the magnetic properties.

\section{Numerical Setup} \label{sec:setup}

The numerical simulation was performed with the radiative MHD code {\it R2D2}, which stands for the Radiation and RSST (reduced speed of sound technique) for Deep Dynamics \citep[see][for the details]{2019SciA....5.2307H}. In brief, this code solves the MHD equations with taking into account the radiative energy transfer and adopts RSST \citep{2005ApJ...622.1320R,2012A&A...539A..30H,2015ApJ...798...51H,2019A&A...622A.157I} to deal with the fast sound speed and mitigate the Courant-Friedrichs-Lewy condition. We adopt the gray approximation for the radiative transfer with the Rosseland mean opacity.

The Cartesian box spans $98.3\ {\rm Mm}\ (x) \times 98.3\ {\rm Mm}\ (y) \times 139.9\ {\rm Mm}\ (z)$, resolved by a $1024\times 1024\times 256$ grid. The horizontal ($x$ and $y$) grid spacing is $96\ {\rm km}$ (uniform), while the vertical ($z$) spacing increases from $48\ {\rm km}$ around the top boundary to $2950\ {\rm km}$ around the bottom boundary. The top boundary is located $700\ {\rm km}$ above the average $\tau=1$ surface. The bottom boundary is thus located at $-139.2\ {\rm Mm}$ (i.e. $0.8R_{\odot}$), which is deep enough compared to the thickness of the convection zone ($\sim 200\ {\rm Mm}$). Therefore, we can self-consistently solve large-scale convection and surface granulation at the same time.

We assume periodic boundaries for both horizontal directions. The top boundary is open for upflows and closed for downflows, while the density and entropy perturbation from the initial state are free there. The bottom is open for the flows. For mass conservation, the horizontally averaged density is fixed to the initial condition, and the perturbation from the average is free. At the bottom, the entropy in upflows is fixed to the initial value and is free in downflows. We adopt the stress-free boundary condition for the horizontal velocity at both the top and bottom boundaries. The magnetic field at the top is matched to a potential field above, whereas all three components of the field are symmetric about the bottom.

We calculated the convection without magnetic field, first for 60 solar days over a domain up to $-2.35\ {\rm Mm}$ ($0.997R_{\odot}$) with an artificial cooling layer and then for five days over the whole domain until a statistically equilibrium state was attained. This procedure is justified because the existence of the surface does not influence the deep convection structure \citep{2019SciA....5.2307H}. Then, at $t=0\ {\rm hr}$, we introduced a horizontal flux tube at $-16.7\ {\rm Mm}$, which is given as an $x$-directed force-free Lundquist field \citep{1951PhRv...83..307L}:
\begin{eqnarray}
  B_{x}=B_{\rm tb}J_{0}(\alpha r),\, B_{\phi}=B_{\rm tb}J_{1}(\alpha r),
\end{eqnarray}
where $r$ is the radial distance from the axis, $B_{\rm tb}=10\ {\rm kG}$ the axial field strength, $J_{0}$ and $J_{1}$ the Bessel functions, $\alpha=a_{0}/R_{\rm tb}$, $a_{0}=2.404825$, and $R_{\rm tb}=7\ {\rm Mm}$ the tube's radius. The total axial flux is $6.6\times10^{21}\ {\rm Mx}$. Unlike many of the previous $\delta$-spot simulations where the tube was artificially made buoyant or kinematically inserted, the tube here was initially in mechanical balance and thus started moving only in response to the background convection flows.

It should be noted that because of the Alfv\'{e}n speed limiting \citep{2009Sci...325..171R}, with which we limit the Alfv\'{e}n speed to a maximum of $40\ {\rm km\ s}^{-1}$ to accelerate the computation, the physical quantities may be affected, especially within the umbral cores in the upper photosphere. 
In the next section, we focus on the generation and properties of $\delta$-spots. However, readers may consult H. Hotta \& H. Iijima (in preparation) for a detailed account of flux emergence and spot formation.

\section{Results} \label{sec:results}

\subsection{General Evolution} \label{subsec:general}

\begin{figure*}
\begin{center}
\includegraphics[width=1\textwidth]{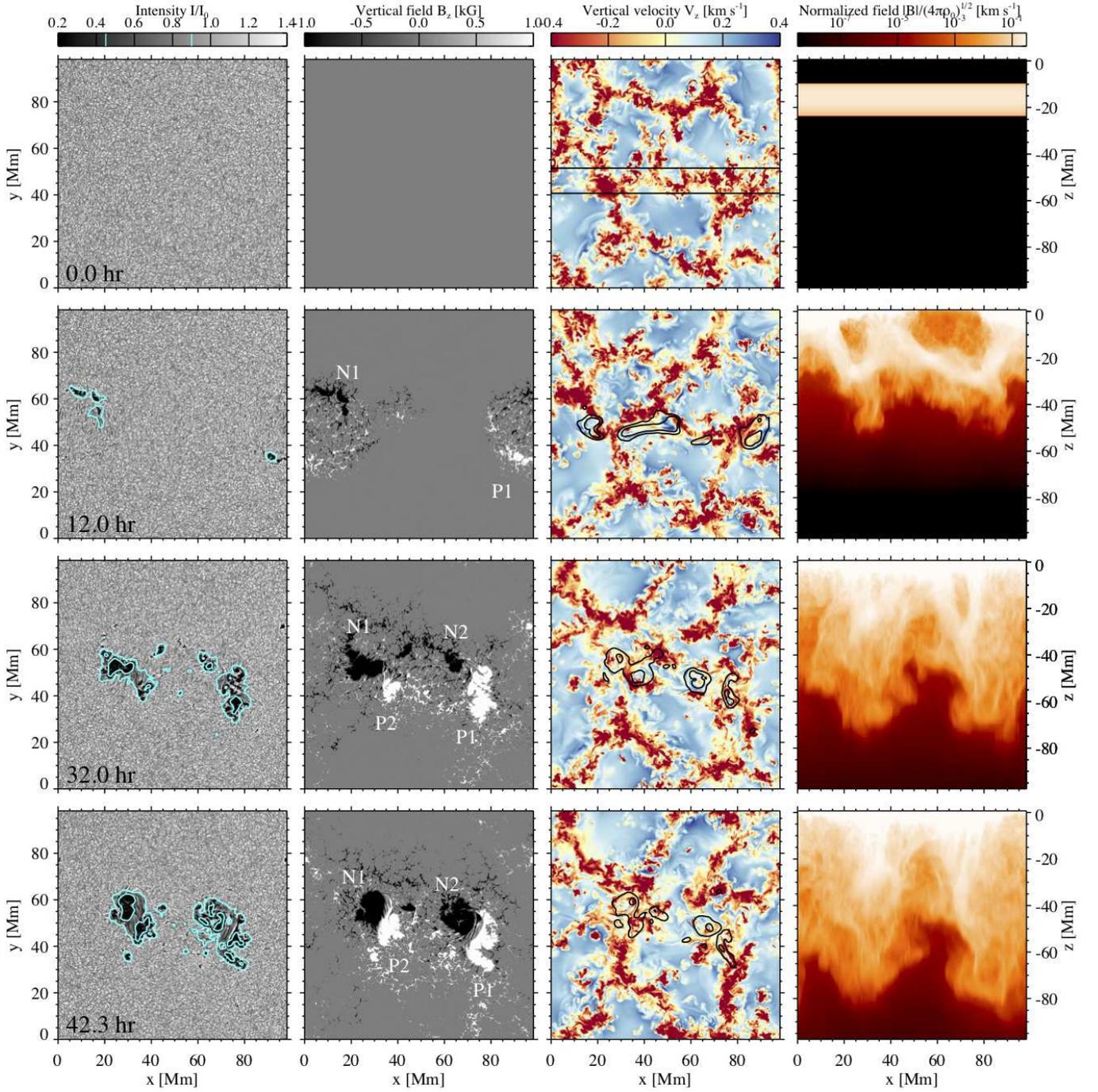}
\end{center}
\caption{Time sequence of (left) the emergent intensity normalized by the quiet-Sun value ($I/I_{0}$), (second) the vertical magnetic field strength ($B_{z}$) sampled at $\tau=1$, (third) the vertical velocity ($V_{z}$) at $-16.7\ {\rm Mm}$ (the initial depth of the tube axis), and (right) the absolute field strength averaged in the $y$-direction and normalized by the local background density (see main text for the details). Turquoise contours indicate where the smoothed intensity is less than $I/I_{0}=0.45$ (umbra) and 0.9 (penumbra), while black contours show $|\mbox{\boldmath $B$}|=5$ and 10 kG. The two emerging bipoles P1-N1 and P2-N2 collide to form the two $\delta$-spots N1-P2 and N2-P1. See the accompanying video for the temporal evolution.\label{fig:general}}
\end{figure*}

Figure \ref{fig:general} shows the temporal evolution of the emerging flux tube. The left and second columns are the emergent intensity normalized by its quiet-Sun average ($I/I_{0}$) and the vertical field strength at the $\tau=1$ surface ($B_{z}$: magnetogram), respectively. The contours indicate the umbra/penumbra and penumbra/quiet-Sun boundaries, which are defined as $I/I_{0}=0.45$ and 0.9, respectively, measured on the intensity map smoothed using a convolution with a Gaussian kernel with a FWHM of 3 Mm \citep{2015ApJ...814..125R}. The third column shows the vertical velocity ($V_{z}$) at $-16.7\ {\rm Mm}$, i.e., the initial depth of the tube axis. The column on the right presents the normalized magnetic field strength, defined as
\begin{eqnarray}
  B'(x,z)=\frac{1}{\Delta y}\int_{y_{1}}^{y_{2}}|\mbox{\boldmath $B$}(x,y,z)|\,dy
  \bigg/ \sqrt{4\pi\rho_{0}(z)},
\end{eqnarray}
where $y_{1}=24.6\ {\rm Mm}$, $y_{2}=73.8\ {\rm Mm}$, $\Delta y=y_{2}-y_{1}$, and $\rho_{0}(z)$ is the initial background density. This quantity indicates the field strength averaged in the $y$-direction and normalized by the square root of the local initial density and thus possesses the dimension of velocity.

At $t=0\ {\rm hr}$, a strong upflow starts to elevate the flux tube around $x=0\ {\rm Mm}$ (i.e. $x=98.3\ {\rm Mm}$ because of the periodic side boundaries) and creates an $\Omega$-shaped loop, which develops into an emerging bipole P1-N1 on the surface (see panels for 12.0 hr). Between P1 and N1, a number of small-scale magnetic elements are scattered and exhibit a net-like structure. Through merging and cancellation, they develop into mature sunspots with penumbrae. These behaviors resemble previous observations and models of flux emergence \citep[e.g.,][]{1996A&A...306..947S,2004ApJ...614.1099P,2003ApJ...582.1206F,2010ApJ...720..233C}.

Meanwhile, in the convection zone, another rising section appears around $x=40\ {\rm Mm}$ (see the $V_{z}$ map for 12.0 hr) and produces the secondary $\Omega$-loop, which eventually appears as the bipole P2-N2 at the domain center between N1 and P1 (see panels for 32.0 hr). The spot area --- measured as the total pixels with $I/I_{0}<0.9$ in the smoothed intensity map --- and the total unsigned flux attain their peak values of $8.0\times 10^{8}\ {\rm km}^{2}$, equivalent to 260 MSH (millionths of the solar hemisphere), and $2.5\times 10^{22}\ {\rm Mx}$, respectively, at 40 hr.

Because the original horizontal tube was elevated at two segments, the legs of the two $\Omega$-loops approach each other to form the spot pairs N1-P2 and N2-P1. As time progresses, each pair collides and eventually shares a common penumbra, building up a $\delta$-spot (see panels for 43.3 hr). Within each $\delta$-spot, elongated dark convection cells are trapped between the umbrae of opposite polarities ($\delta$-spot light bridge). In the accompanying movie, the spots show vigorous rotational motions in the counter-clockwise direction and the spots are connected below the photosphere by U-loops that are stretched down to about $-40\ {\rm Mm}$ by strong downdrafts.

\subsection{Sunspot Rotation} \label{subsec:spotrot}

\begin{figure*}
\begin{center}
\includegraphics[width=\textwidth]{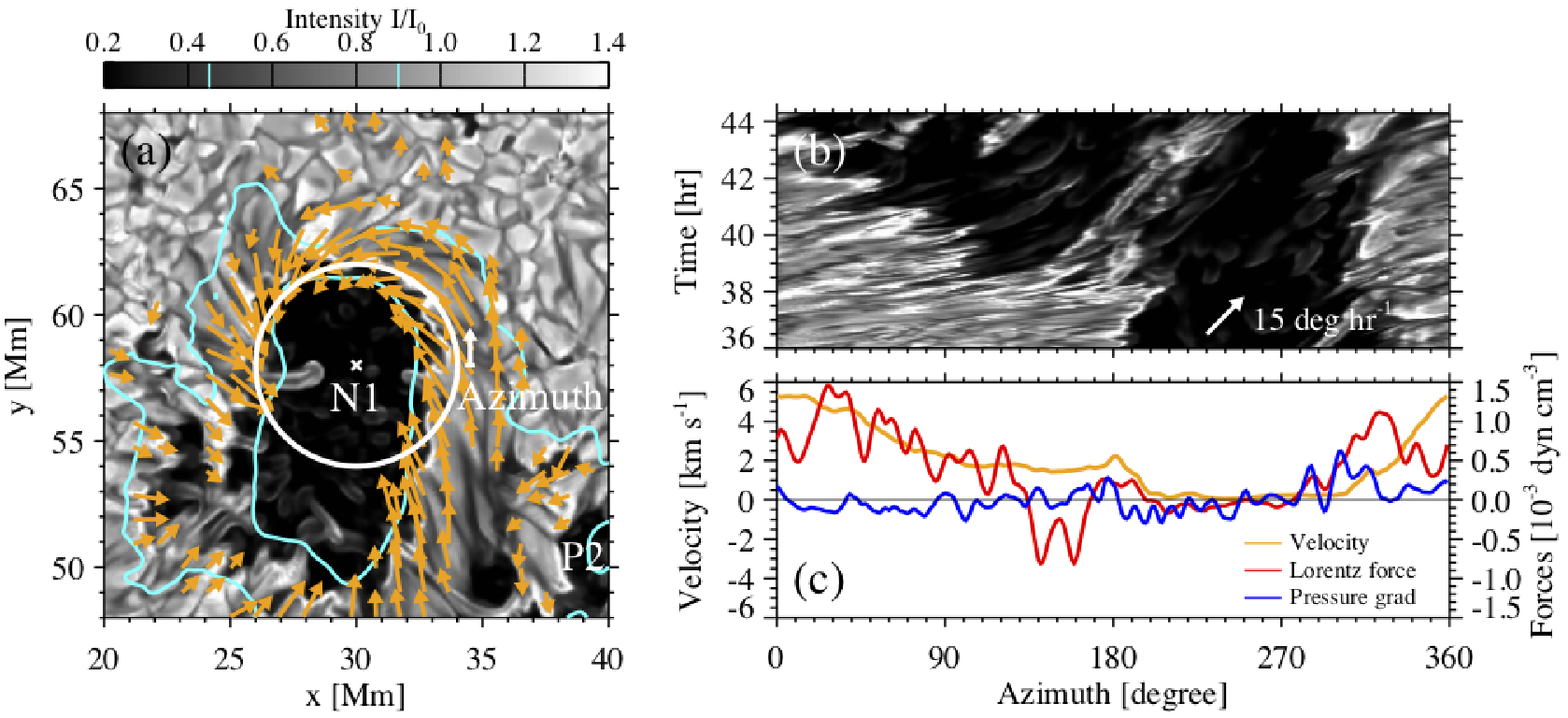}
\end{center}
\caption{(a) Horizontal velocity vectors $(V_{x}, V_{y})$ averaged over 8.3 hr from $t=36.0\ {\rm hr}$, plotted on the intensity map at $t=40.0\ {\rm hr}$. The definition of the turquoise contours is the same as in Figure \ref{fig:general}. (b) Time slice of the intensity map along the circular slit in panel (a), which is centered at $(x, y)=(30\ {\rm Mm}, 58\ {\rm Mm})$ with the radius of 4 Mm. (c) Azimuthal components of the horizontal velocity (orange), the Lorentz force (red), and the gas pressure gradient (blue) along the slit, all averaged over 8.3 hr.\label{fig:spotrot}}
\end{figure*}

In order to examine the driving mechanism of the observed spot rotations, we focus our attention on N1 in Figure \ref{fig:spotrot}. The velocity vectors averaged over 8.3 hr in panel (a) demonstrate that the rotation is not the apparent effect but the actual plasma motion. One may find that the convection cells in the penumbra and light bridge are stretched in the direction of the spot rotation. Panel (b) shows the temporal evolution of the spot structure along the circular slit around N1, which clearly shows an apparent counter-clockwise motion with the angular velocity up to $15^{\circ}\ {\rm hr}^{-1}$.

The azimuthal component of the temporally averaged plasma velocity along the slit is plotted in panel (c). The velocity ranges from almost $0\ {\rm km\ s}^{-1}$ in the umbral regions to more than $5\ {\rm km\ s}^{-1}$ in the penumbra and the light bridge. Overplotted are the azimuthal components of the Lorentz force ($(\nabla\times\mbox{\boldmath $B$})\times\mbox{\boldmath $B$}/(4\pi)$) and gas pressure gradient ($-\nabla p$) along the slit. In most of the regions, the Lorentz force dominates the pressure gradient, indicating that the main driving force of the spot rotation is the Lorentz force.

The above result is akin to those of the previous flux emergence simulations and is explained in the following manner \citep[e.g.,][]{2000ApJ...545.1089L,2009ApJ...697.1529F,2015A&A...582A..76S}. When an isolated flux tube with a left-handed twist emerges into the low-density atmosphere, it drastically expands and the field lines are therefore bent in a manner that the Lorentz force acts in the direction opposite to the original twist, i.e., counter-clockwise. The observed counter-clockwise spot rotations are a consequence of this Lorentz force, and the flux tube injects its helicity into the atmosphere through this process.

\subsection{Magnetic Properties of the $\delta$-spot PIL} \label{subsec:pils}

\begin{figure*}
\begin{center}
\includegraphics[width=0.95\textwidth]{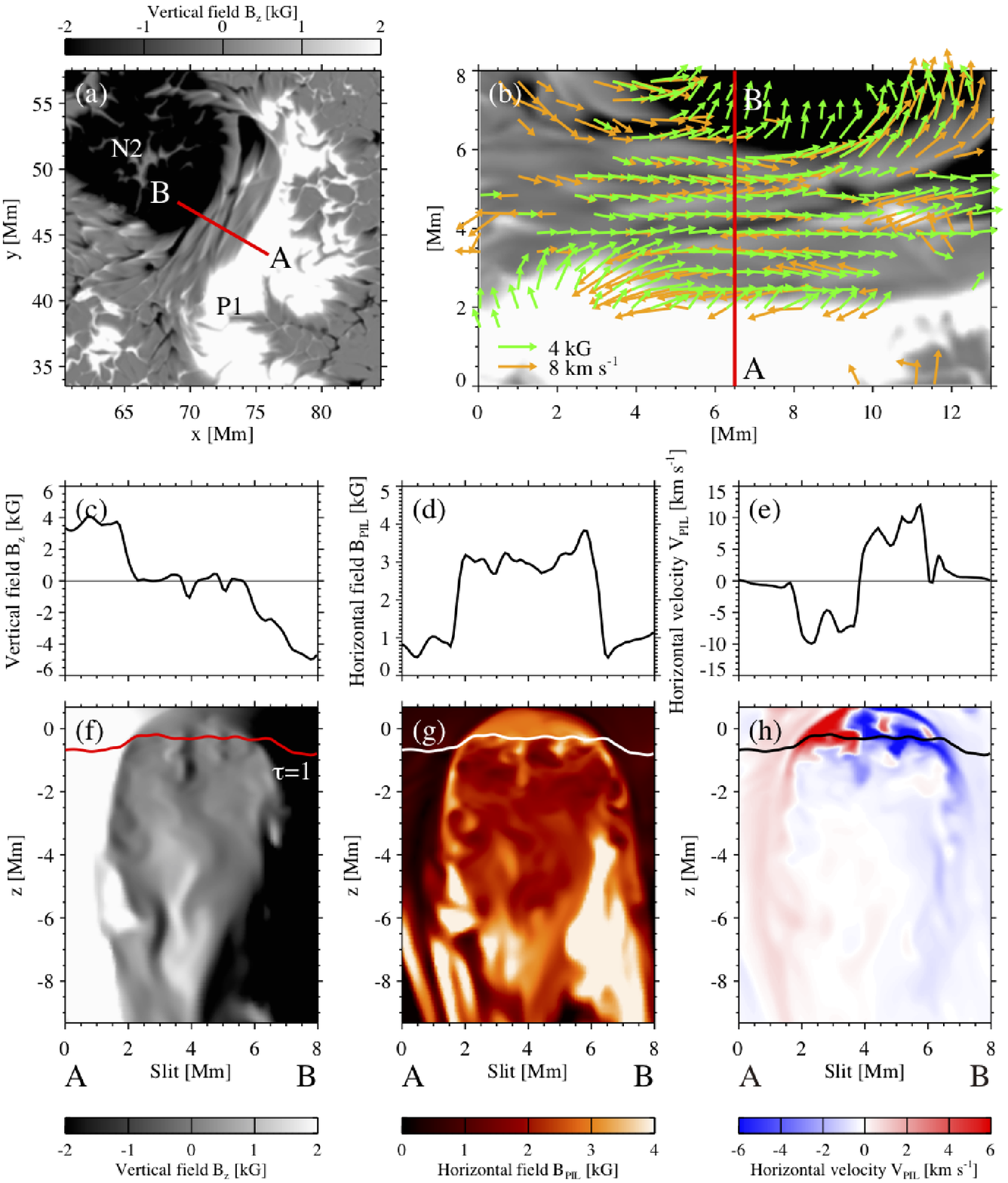}
\end{center}
\caption{(a) Magnetogram at $t=42.3\ {\rm hr}$ with the slit A-B (red). (b) Enlarged magnetogram around the slit with arrows indicating the horizontal magnetic field (green) and the horizontal velocity (orange). (c--e) Variations along the slit A-B of the vertical field ($B_{z}$) and the horizontal magnetic and velocity fields along the PIL, sampled at $\tau=1$ ($B_{\rm PIL}$ and $V_{\rm PIL}$). (f--h) Their cross-sectional profiles, on which the $\tau=1$ layer is plotted as a solid curve.\label{fig:pil}}
\end{figure*}

Figure \ref{fig:pil} reveals the detailed magnetic and velocity structures of the PIL in the $\delta$-spot N2-P1. As in panel (a), the PIL shows an alternating pattern of elongated positive and negative polarities (magnetic channel). Panel (b) demonstrates that a strong counter-streaming flow of up to $10\ {\rm km\ s}^{-1}$ occurs between the two rotating umbrae. The flow runs along the PIL and the horizontal field vectors are aligned with the flow vectors in most parts, constituting a highly-sheared PIL.

Cross-sectional profiles in panels (c--h) reveal that the $\tau=1$ surface is elevated in the PIL by a few 100 km. The vertical magnetic field inverts its sign across the PIL with a steep gradient of about $1\ {\rm kG\ Mm}^{-1}$ on average. The horizontal magnetic field is strongly sheared and intensified by the horizontal flows, with the field strength being up to about 4 kG. One may find that the strong $B_{z}$ concentrations in the magnetic channel structure appear at the locations of steep velocity shear.

As shown above, the velocity shear of the counter-streaming flow, driven by the Lorentz force acting on each spot umbra (Section \ref{subsec:spotrot}), intensifies the magnetic shear at the PIL. In the previous ideal $\delta$-spot model by \citet{2017ApJ...850...39T} that lacks thermal convection, the horizontal field at the PIL was in fact intensified, but only up to an equipartition field strength of about 1 kG. In the present model, on the contrary, the vigorous convection continues around the PIL ($\delta$-spot light bridge), which further strengthens the field to 4 kG.

\begin{figure}
\begin{center}
\includegraphics[width=0.45\textwidth]{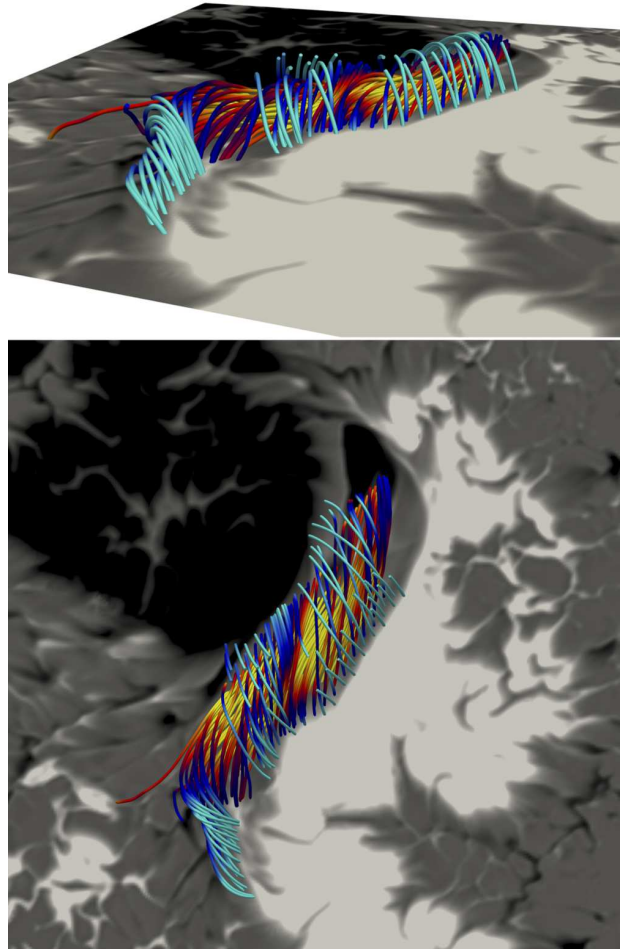}
\end{center}
\caption{Side and top views of magnetic field lines above the PIL of the spot N2-P1 at $t=42.3\ {\rm hr}$. The color of the field lines represents the strength of $B_{x}$, in which $B_{x}>0$ ($<0$) is colored yellow to red (aqua to blue). For better visualization, the vertical scale is stretched by a factor of two. The bottom plane is the magnetogram with the same field of view as Figure \ref{fig:pil}(a).\label{fig:3d}}
\end{figure}

The magnetic structure above the surface is presented in Figure \ref{fig:3d}. Above the PIL, the field lines are low-lying and highly sheared (yellow to red), which is covered by the less-sheared field lines that extend from the umbrae on both sides (aqua to blue). The overall field configuration takes the form of a twisted flux rope, which is often observed in the corona in the pre-eruption phase. The sheared field at the flux rope core is due to the advection by the sheared flows, whereas the overlying arcade is more potential because the umbral cores rotate less strongly (see Figures \ref{fig:spotrot}(a) and \ref{fig:pil}(b)).


\section{Discussion} \label{sec:discussion}

In this Letter, we reported on the spontaneous generation of $\delta$-spots, which are known to be flare-productive, by performing a realistic flux emergence simulation. Thanks to the deep enough computation domain realized by the state-of-the-art R2D2 code, we can assess the effects of radiative transfer and thermal convection on emerging flux. Although some physical quantities may not be directly compared with observations because of the numerical assumptions, the general properties revealed in this study lead to a better understanding of the genesis of flare-productive regions.

The initial force-free flux tube was pushed up by large-scale upflows at two separate portions as a pair of emerging bipoles. In the photosphere, umbrae of opposite polarities collided with each other and formed a pair of strongly-packed $\delta$-spots because their legs were connected by deeply-anchored U-loops.\footnote{Needless to say, it is because we applied periodic side boundaries that we observed {\it two} $\delta$-spots.} This result points to the possibility that the multi-buoyant segment scenario occurs on the actual Sun.

One interesting hypothesis we can derive from our results is that the $\delta$-spot formation, and therefore the resultant flare eruptions, might be a probabilistically determined process depending on where a magnetic flux is located on the turbulent background. In fact, in some of the test cases where we only changed the location of the initial flux tube in the same background convection, we found that the tube appears as a single emerging bipole and never produces $\delta$-spots (H. Hotta \& H. Iijima, in preparation).

Another point to be discussed is the effect of deep convection structures. Helioseismology points out that convections in numerical simulations are prone to deviation from the observations, especially in large scales ($>30\ {\rm Mm}$) \citep{2012PNAS..10911928H,2014ApJ...793...24L}. However, because in the present model, we selected the small horizontal domain extent of $\sim 100\ {\rm Mm}$, which inhibits the generation of unreasonably large cells, the cell sizes of $>30\ {\rm Mm}$ may not be critically different from the reality. Also, the spot generation may be impacted by the choice of the bottom boundary depth and thus the convective patterns in deep layers. However, again, since the present $\delta$-spots stem from the supergranule-scale emerging flux, even if we limit the bottom depth to, say, $-30\ {\rm Mm}$, the result may not differ much as far as we keep the domain that is deep enough to harbor supergranulation.

Each polarity of the generated $\delta$-spots showed a strong rotation for hours, which was driven by the Lorentz force via unwinding of the tube's twist. The observed angular speed of $15^{\circ}\ {\rm hr}^{-1}$ is somewhat larger than the reported values \citep[e.g.,][]{2009SoPh..258..203M}. The rotation speed may depend on the tube's initial twist, but the potential-field top boundary may allow for the rapid helicity release as well \citep{2010ApJ...720..233C}.

The collision of two rotating spots of opposite polarity generated a counter-streaming flow at the PIL, which greatly enhanced the magnetic shear and horizontal field. The observed field strength of 4 kG was due to the velocity shear as well as the light bridge convection. A magnetic channel, some of which is suggested to be a flare-triggering field \citep[e.g.,][]{2012ApJ...760...31K}, was created by the strong velocity shear, implying that small-scale local convection may play a crucial role in evoking flare eruptions \citep{2013ApJ...773..128T}.

Magnetoconvective property of a $\delta$-spot light bridge is in many ways similar to that of a regular light bridge that separates the umbrae of the same polarity. Comparing the cross-sectional profiles in Figure \ref{fig:pil} of this Letter and Figure 4 of \citet{2015ApJ...811..138T}, one may find that in the regular bridge, the horizontal flow is dominant and the elevated iso-$\tau$ surfaces show a cusp structure that is sandwiched by the canopy field fanning out from the adjacent umbrae, whereas in the $\delta$-spot bridge, a counter-streaming flow is confined by the low-lying sheared field that connects the neighboring umbrae.

The objective of this Letter is to reveal how the turbulent solar convection allows for the spontaneous formation of $\delta$-spots, sheared PILs, magnetic channels, and twisted flux ropes, all of which are the key ingredients of flare-productive active regions.
In the forthcoming papers, we intend to perform detailed analyses on the generation of sunspots, strong PIL horizontal fields, and coronal response to the $\delta$-spot formation.

\acknowledgments

The authors are grateful to the anonymous referee for improving the manuscript.
The results were obtained using the K computer at the RIKEN (proposal nos. hp190070, hp180215, hp180042, and ra000008).
This work was supported by JSPS KAKENHI Grant Numbers JP15H05814 (PI: K. Ichimoto) and JP18H04436 (PI: H. Hotta), by the NINS program for cross-disciplinary study (Grant Numbers 01321802 and 01311904) on Turbulence, Transport, and Heating Dynamics in Laboratory and Astrophysical Plasmas: ``SoLaBo-X'', and by MEXT as Exploratory Challenge on Post-K Computer (Elucidation of the Birth of Exoplanets [Second Earth] and the Environmental Variations of Planets in the Solar System).

%

\vspace{5mm}
\facilities{K computer}


\software{R2D2 \citep{2019SciA....5.2307H}}






\bibliography{toriumi2019}{}
\bibliographystyle{aasjournal}



\end{document}